# Electronic band structure and Fermi surface for new layered superconductor $LaO_{0.5}F_{0.5}BiS_2$ in comparison with parent phase $LaOBiS_2$ from first principles

*I.R. Shein,\* A.L. Ivanovskii*

*Institute of Solid State Chemistry, Ural Branch of the Russian Academy of Sciences, 620990, Ekaterinburg, Russia*

*\* E-mail: shein@ihim.uran.ru*

By means of first-principles calculations, we have probed the peculiarities of the electronic band structure and Fermi surface for the recently discovered layered superconductor $LaO_{0.5}F_{0.5}BiS_2$ in comparison with the parent phase $LaOBiS_2$. The electronic factors promoting the transition of $LaOBiS_2$ upon fluorine doping to superconducting state: inter-layer charge transfer, the evolution of the Fermi surface, and the dependence of the near-Fermi densities of states on x for $LaO_{1-x}F_xBiS_2$ are evaluated and discussed in comparison with the available experiments.

The superconducting materials with layered structures attract currently much attention owing to a set of intriguing physical properties related to unconventional mechanisms of superconductivity. The most popular materials among them are now Fe-pnictogen (*Pn*) and Fe-chalcogen (*Ch*) based compounds, which form the groups of so-called 11, 111, 122, 1111, 23335, 42226, and more complex phases, reviews [1-7]. These materials are composed of superconducting blocks [Fe*Pn*] or [Fe*Ch*], which are separated by so-called spacer blocks. In all these systems, which possess remarkable chemical flexibility, success in increasing the critical temperature $T_C$ was achieved mainly by optimization of the structural parameters and (or) *via* changes in the carrier doping levels.

The very recent discovery of superconductivity in $Bi_4O_4S_3$ ($T_C$ ~ 4.5 K) [8-11] and then in several related $BiS_2$-based systems ($LaO_{1-x}F_xBiS_2$, $CeO_{1-x}F_xBiS_2$, $PrO_{1-x}F_xBiS_2$, and $NdO_{1-x}F_xBiS_2$, with $T_C$ ~ 10.0 K [12], 3.0 K [13], 5.5 K [14], and 5.6 K [15], respectively), which contain identical blocks [BiS$_2$], has opened the next stage in the field of modern layered superconductors and provided a new research domain of physics for such systems.

It is remarkable is that among all BiS$_2$-based superconductors (SCs) discovered so far, the maximum $T_C$ (~10 K) is achieved for La(O,F)BiS$_2$, *i.e.* at electron doping of the parent semiconducting phase LaOBiS$_2$ *via* partial replacement of oxygen by fluorine [12,16-18]. Thus, these newest layered BiS$_2$-based SCs, like the aforementioned Fe-(*Pn,Ch*) systems, are very sensitive to the carrier doping level, when atomic substitutions cause profound changes in their properties, and superconductivity appears in the vicinity of the insulating-like state. Note also that a non-monotonic change of $T_C$ was achieved by doping of the oxide blocks [LaO], which act as spacer blocks, when for LaO$_{1-x}$F$_x$BiS$_2$ superconductivity appears at x = 0.2, grows with further F-doping, and reaches $T_C$ ~10 K at x = 0.5, and then it is gradually suppressed and disappears at x = 0.7.

From the theoretical standpoint, for the family of BiS$_2$-based SCs, the available band structure calculations [8,19-21] reveal that for Bi$_4$O$_4$S$_3$ in the window around the Fermi level ($E_F$) the main contributions come from the Bi $6p_{x,y}$ orbitals, which are the most active pairing states [8]; on the other hand, LaOBiS$_2$ belongs to band insulators [16,19]. Using the band structure of LaOBiS$_2$, H. Usui *et al.* [19] have constructed a minimal electronic model for superconducting LaO$_{1-x}$F$_x$BiS$_2$, and declared a good nesting of the Fermi surface (FS) at around x ~ 0.5. Very recently, *Li and Xing* [20] have calculated the electron-phonon coupling constant for LaO$_{0.5}$F$_{0.5}$BiS$_2$ ($\lambda$ = 0.8) and concluded that this material is a conventional electron-phonon SC. On the other hand, *Yildirim* [21] found that LaO$_{0.5}$F$_{0.5}$BiS$_2$ possesses rather unusual structural and dynamical properties and is a strong electron-phonon coupled SC in the vicinity of competing ferroelectric and charge density wave (CDW) phases, when large-amplitude in-plane displacements of sulfur atoms control the structural properties and give rise to large electron-phonon coupling.

In view of these circumstances, in this Letter we present detailed calculations of the band structure and FS for the newest SC LaO$_{0.5}$F$_{0.5}$BiS$_2$ - in comparison with the parent phase LaOBiS$_2$. Besides, using the obtained electronic structure of LaO$_{0.5}$F$_{0.5}$BiS$_2$ we discuss the changes in the near-Fermi density of states (DOSs) and in the FS topology for LaO$_{1-x}$F$_x$BiS$_2$ in the doping interval 0.2 ≤ x ≤ 0.7 - in comparison with available experiments.

We began with the parent tetragonal phase LaOBiS$_2$, which adopts the space group *P*4/*nmm* with lattice parameters *a* = 4.05 Å and *c* =13.74 Å [19]. This phase possesses a layered structure composed of blocks [LnO] and [BiS$_2$], which are stacked in the sequence …[LnO]/[BiS$_2$]/[LnO]/[BiS$_2$]/… as depicted in Fig. 1. After that we considered the fluorine-doped system with the formal stoichiometry LaO$_{0.5}$F$_{0.5}$BiS$_2$, for which the maximal $T_C$ was found [12,16-18]. Here the experimentally obtained [12] lattice parameters (*a* = 4.0527 Å and *c* = 13.3237 Å) and atomic coordinates La: (0.5, 0, 0.1015); Bi: (0.5, 0, 0.6231); S$_1$: (0.5, 0, 0.3657); S$_2$: (0.5, 0, 0.8198); O/F (with occupancy 0.5/0.5): (0, 0, 0), were used. Note that the substitution of F$^{-1}$ for O$^{-2}$ leads to an *anisotropic* deformation of the lattice, when the parameter *a* remains almost invariable (in comparison with the parent phase LaOBiS$_2$), whereas the parameter *c* becomes compressed by 3 %; and this situation is typical of layered phases at atomic substitutions, reviews [1-7].

Our calculations were performed by means of the full-potential method within mixed basis APW+lo (FLAPW) implemented in the WIEN2k suite of programs [22]. The generalized gradient approximation (GGA) to exchange-correlation potential [23] was used. The densities of states (DOSs) were obtained by the modified tetrahedron method [24].

The band structure, the total, atomic, and orbital decomposed partial densities of states for the parent phase $LaOBiS_2$ are depicted in Figs. 2 and 3. We see that the valence band (VB, in the energy interval from -5.6 eV to the Fermi level) consists mainly of the $p$ states of O and S atoms; the contributions of the valence orbitals of La and Bi are much smaller. The VB is separated from the conduction band (CB) by a gap indicating an insulating behavior of $LaOBiS_2$. The minimal indirect gap ($Z$ - $R$ transition) is estimated at about 0.4 eV. Note that the bottom of the CB for $LaOBiS_2$ is formed mainly by unoccupied Bi $6p$ states with admixtures of S $3p$ states; thus at electron doping, these states are expected to be filled.

Our calculations reveal that for $LaO_{0.5}F_{0.5}BiS_2$, in comparison with $LaOBiS_2$, the Fermi level is displaced upwards and intersects six bands along $Z$-$R$, $X$-$M$ and $A$-$Z$ high-symmetry directions of the Brillouin zone (BZ), which form the electronic pockets of the FS. Thus, partial substitution of fluorine for oxygen leads to a metallic-like behavior of the cubic $LaO_{0.5}F_{0.5}BiS_2$, where the carriers are electrons, see Figs. 2 and 3.

Generally, it is quite an expected result, since for oxides, fluorine with one additional electron compared to oxygen may be considered as a "universal" electron dopant, which, owing to injection of additional electrons, changes the insulating oxides to metallic conductors or superconductors. For example, superconductivity was found at fluorine doping in simple insulating oxides such as $WO_3$, see [25,26].

However for the discussed $LaO_{0.5}F_{0.5}BiS_2$ the situation became more complicated: as can be seen from the DOSs picture (Fig. 3), though fluorine doping occurs in the spacer blocks [LnO], the main effect of metallization takes place for the neighbor blocks [BiS$_2$], which are involved in the formation of the superconducting state. At the same time, the semiconducting spectrum of the oxide blocks (for $LaOBiS_2$) is transformed for $LaO_{0.5}F_{0.5}BiS_2$ into a spectrum typical of so-called "bad metals", with very low densities of states at the Fermi level.

This effect may be explained taking into account the charge distributions between the adjacent blocks: [LnO]/[BiS$_2$]. For $LaOBiS_2$, assuming the usual oxidation numbers of atoms: $La^{3+}$, $Bi^{3+}$, $O^{2-}$, and $S^{3-}$, the simple ionic formula should be: $(La^{3+}O^{2-})^{1+}(Bi^{3+}S^{2-}_2)^{1-}$; this implies that the charge transfer between the adjacent blocks is equal: [LnO] →[BiS$_2$] = 1$e$, and the blocks [LnO] act as "charge reservoirs". To estimate the effective atomic charges ($Q^{eff}$) numerically, we carried out a Bader [27] analysis. For $LaOBiS_2$, the values of $Q^{eff}$ are: $(La^{1.990+}O^{1.271-})^{0.719+}(Bi^{2.020+}S_{(1)}^{1.022-}S_{(2)}^{1.717-})^{0.719-}$ (here $S_{(1)}$ and $S_{(2)}$ denote the non-equivalent S atoms in blocks [BiS$_2$], see [12,16-18]) and the charge transfer [LnO] →[BiS$_2$] is about 0.7$e$. The reduction of the effective charges (in comparison with the ideal ionic model) is due to the overlapping of the valence states and to the formation of covalent bonds, see Fig. 3.

For $LaO_{0.5}F_{0.5}BiS_2$, the values of $Q^{eff}$ are: $(La^{2.012+}O^{1.278-}_{0.5}F^{0.794-}_{0.5})^{0.976+}(Bi^{1.728+}S_{(1)}^{1.099-}S_{(2)}^{1.606-})^{0.976-}$, i.e. fluorine doping of the parent phase increases the charge transfer to the blocks [BiS$_2$] by about 0.26 $e$, promoting the aforementioned primary metallization of these blocks.

Coming back to the electronic properties of $LaO_{0.5}F_{0.5}BiS_2$, let us discuss in greater detail the near-Fermi bands, which are involved in the formation of the superconducting state. From the atomic and orbital decomposed DOSs depicted in Fig. 3 it is seen that the main contribution to this region comes from the Bi $6p$ states, with additions of the S $3p$ states. Numerically, the contributions to the total DOS at the Fermi level, $N^{tot}(E_F)$, are:

$N^{Bi6p}(E_F) = 0.423$ 1/eV·form. unit and $N^{S3p}(E_F) = 0.2087$ 1/eV·form. unit. In turn, for the Bi 6p states, the main contributions are from the in-plane Bi $6p_{x,y}$ orbitals within the blocks [BiS$_2$]: $N^{Bi6p}_{x}(E_F) = 0.206$ 1/eV·form. unit, $N^{Bi6p}_{y}(E_F) = 0.207$ 1/eV·form. unit – in comparison with out-of-plane Bi $6p_z$ orbitals: $N^{Bi6p}_{z}(E_F) = 0.010$ 1/eV·form. unit. The calculated value of $N^{tot}(E_F)$ for LaO$_{0.5}$F$_{0.5}$BiS$_2$ allows us to estimate the Sommerfeld constant ($\gamma$) under the assumption of the free electron model as $\gamma = (\pi^2/3)N(E_F)k^2_B$. The calculated $\gamma = 3.3$ mJ·K$^{-2}$·mol$^{-1}$ agrees reasonably with earlier estimations: $\gamma = 3.0$ mJ·K$^{-2}$·mol$^{-1}$ [20].

The authors [18] noticed that the dome-shaped dependence of $T_C$ on x for LaO$_{1-x}$F$_x$BiS$_2$ (with maximal $T_C \sim 10$ K for x=0.5) resembles the x dependence of the lattice parameter *a*, therefore the structural factor should be important for superconductivity in these materials.

On the other hand, to evaluate the role of the electronic factor, we probed the x dependence of $N^{tot}(E_F)$. For this purpose, we varied the electronic concentration, simulating (in the framework of the rigid band model) the doping level in LaO$_{1-x}$F$_x$BiS$_2$ in the range 0.2 < x < 0.7.

The calculated changes in $N^{tot}(E_F)$ are plotted in Fig. 4 together with the experimental phase diagram of LaO$_{1-x}$F$_x$BiS$_2$ [18]. We see quite a similar dome-shaped dependence of $N^{tot}(E_F)$ on x with the maximal $N^{tot}(E_F)$ for LaO$_{0.5}$F$_{0.5}$BiS$_2$. Thus our calculations show that together with the structural factor [18], the electronic factor provides the transition of LaOBiS$_2$ into the superconducting state upon fluorine doping, and it should be responsible for the regulation of the observed x dependence of $T_C$ for LaO$_{1-x}$F$_x$BiS$_2$.

Finally, let us discuss the Fermi surface for the superconducting LaO$_{0.5}$F$_{0.5}$BiS$_2$ and the evolution of its topology depending on the doping level. For LaO$_{0.5}$F$_{0.5}$BiS$_2$, the Fermi surface adopts a 2D-like topology (which is typical of layered SCs, reviews [1-7]) and comprises a set of disconnected cylinders extended along the $k_z$ direction, Fig. 1. Two of them are along the *Γ-Z* direction, the other two are in the corners of the BZ (along the *M-A* direction), and the rest two small cylinder-like sheets are on the lateral faces of the BZ (along the *X-R* direction). Next, in Fig. 5 we present a set of FSs for LaO$_{1-x}$F$_x$BiS$_2$ across the doping level. We see that in all the cases the FSs retain the quasi-two-dimensional type, but their shapes change essentially depending on the band occupancy (doping level). So at an initial stage of doping (x=0.2), the FS contains only two electron-like sheets in the BZ lateral sides. With an increase in the band occupancy, the volume of the Fermi surface gradually increases, but the shape is retained (x=0.3). Visible reconstruction of the FS occurs for x ~ 0.4, when the disconnected sheets on the lateral sides of the BZ became merged, and new small sheets (along *X-R*) appear. At the highest doping level (x ≥ 0.5), the next stage of reconstruction occurs, when the FS adopts the form described earlier for LaO$_{0.5}$F$_{0.5}$BiS$_2$.

In summary, using first-principles calculations we examined the electronic band structure and the Fermi surface for the very recently discovered superconductor LaO$_{0.5}$F$_{0.5}$BiS$_2$ in comparison with the parent band insulator LaOBiS$_2$ and determined the key electronic factors promoting the superconducting transition for the fluorine-doped LaOBiS$_2$ and the $T_C$ dependence on x for LaO$_{1-x}$F$_x$BiS$_2$. According to our data, the main factors at fluorine doping are: (i) the increase of the charge transfer between adjacent blocks [LnO]→[BiS$_2$] related to metallization of superconducting blocks [BiS$_2$]; (ii) the dome-shaped dependence of $N^{tot}(E_F)$ on x related to the regulation of the values of $T_C$, and (iii) the reconstruction of the 2D-like Fermi surface, which is related to the nesting effect.

———————————————

1. M. V. Sadovskii, Phys.-Usp. **51**, 1201 (2008).
2. A. L. Ivanovskii, Phys.-Usp. **51**, 1229 (2008).
3. F. Ronning, E.D. Bauer, T. Park, et al. Physica C **469**, 396 (2009).
4. Z. A. Ren, and Z. X. Zhao, Adv. Mater. **21**, 4584 (2009).
5. D. C. Johnson, Adv. Phys. **59**, 803 (2010).
6. A.L. Ivanovskii, Russ Chem. Rev. **79**, 1 (2011).
7. A. L. Ivanovskii, Physica C **471**, 409 (2011).
8. Y. Mizuguchi, H. Fujihisa, Y. Gotoh, et al., arXiv:1207.3145 (2012).
9. S. Li, H. Yang, J. Tao, X. Ding, and H.-H. Wen, arXiv:1207.4955 (2012).
10. S. G. Tan, L. J. Li, Y. Liu, P. et al., arXiv:1207.5395 (2012).
11. S. K. Singh, A. Kumar, B. Gahtori, et al., arXiv:1207.5428 (2012).
12. Y. Mizuguchi, S. Demura, K. Deguchi, et al., arXiv:1207.3558 (2012).
13. J. Xing, S. Li, X. Ding, H. Yang, and H. H. Wen, arXiv:1208.5000 (2012).
14. R. Jha, S. K. Singh and V. P. S. Awana, arXiv:1208.5873 (2012).
15. S. Demura, Y. Mizuguchi, K. Deguchi, et al., arXiv:1207.5248 (2012).
16. V. P. S. Awana, A. Kumar, R. Jha, et al., arXiv:1207.6845 (2012).
17. H. Kotegawa, Y. Tomita, H. Tou, et al., arXiv:1207.6935 (2012).
18. K. Deguchi, Y. Mizuguchi, S. Demura, et al., arXiv:1209.3846 (2012).
19. H. Usui, K. Suzuki, and K. Kuroki, arXiv:1207.3888 (2012).
20. B. Li, and Z. W. Xing, arXiv:1210.1743 (2012).
21. T. Yildirim, arXiv:1210.2418 (2012).
22. P. Blaha, K. Schwarz, G.K.H. Madsen, et al., *WIEN2k, An Augmented Plane Wave Plus Local Orbitals Program for Calculating Crystal Properties*, Vienna University of Technology, Vienna, 2001.
23. J. P. Perdew, S. Burke, and M. Ernzerhof, Phys. Rev. Lett. **77**, 3865 (1996).
24. P.E. Blöchl, O. Jepsen, O.K. Andersen, Phys. Rev. B **49**, 16223 (1994).
25. D. Hirai, E. Climent-Pascual and R. J. Cava, Phys. Rev. B **84**, 174519 (2011).
26. I. R. Shein, and A. L. Ivanovskii, JETP Lett., **95**, 66 (2012).
27. R.F.W. Bader, *Atoms in Molecules: A Quantum Theory*, Intern. Series of Monographs on Chemistry, Clarendon Press, Oxford, 1990.

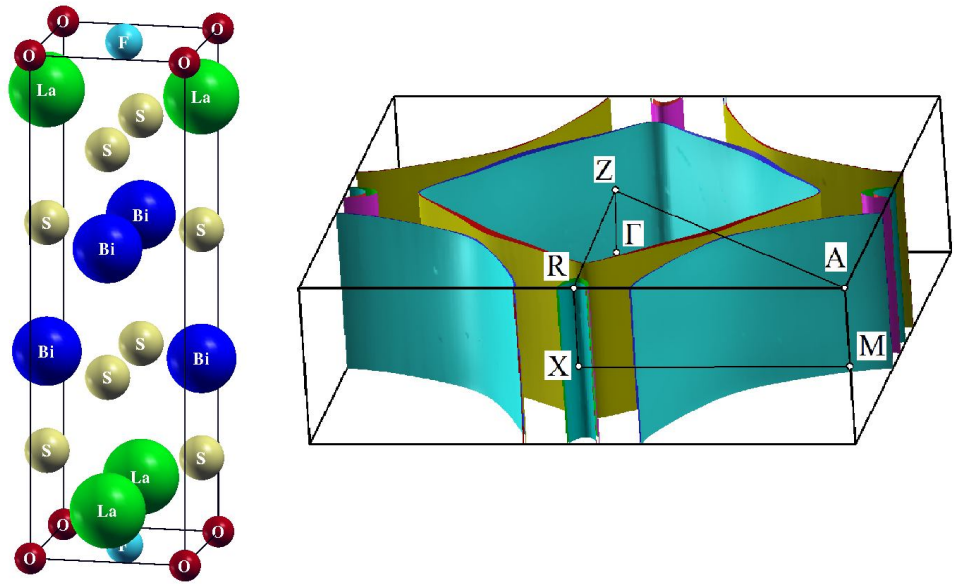

Fig. 1. Crystal structure and calculated Fermi surface for $LaO_{0.5}F_{0.5}BiS_2$

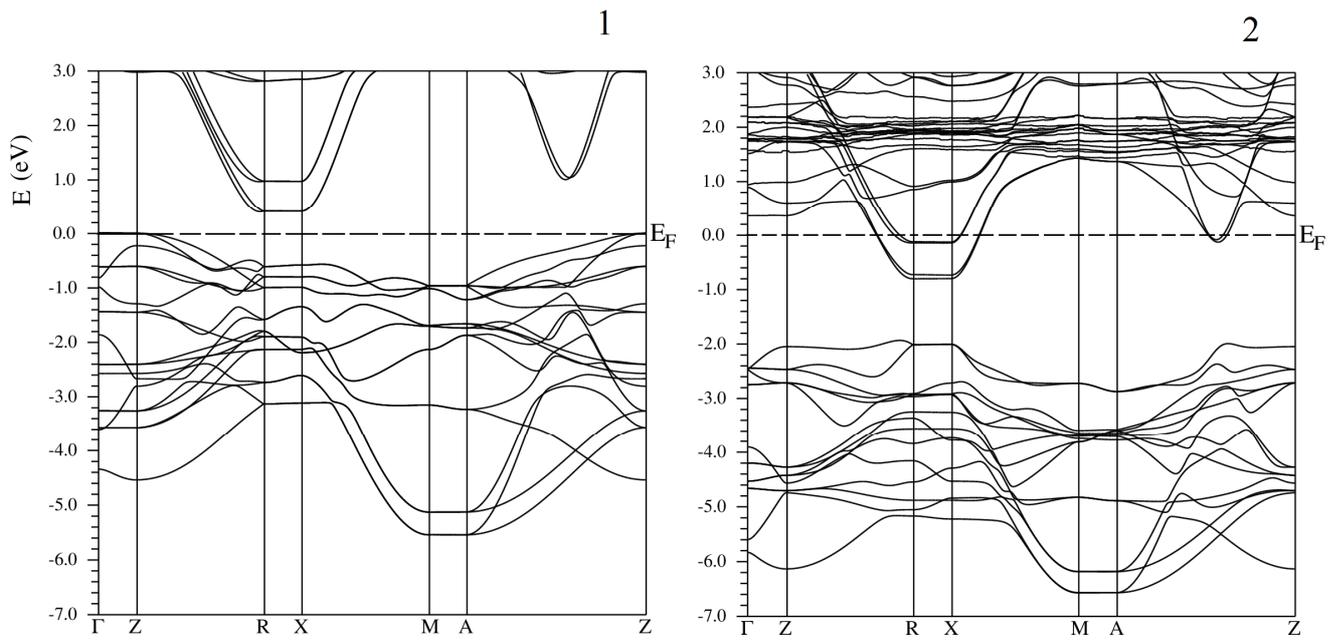

Fig. 2. Electronic bands for $LaOBiS_2$ (1) and $LaO_{0.5}F_{0.5}BiS_2$ (2).

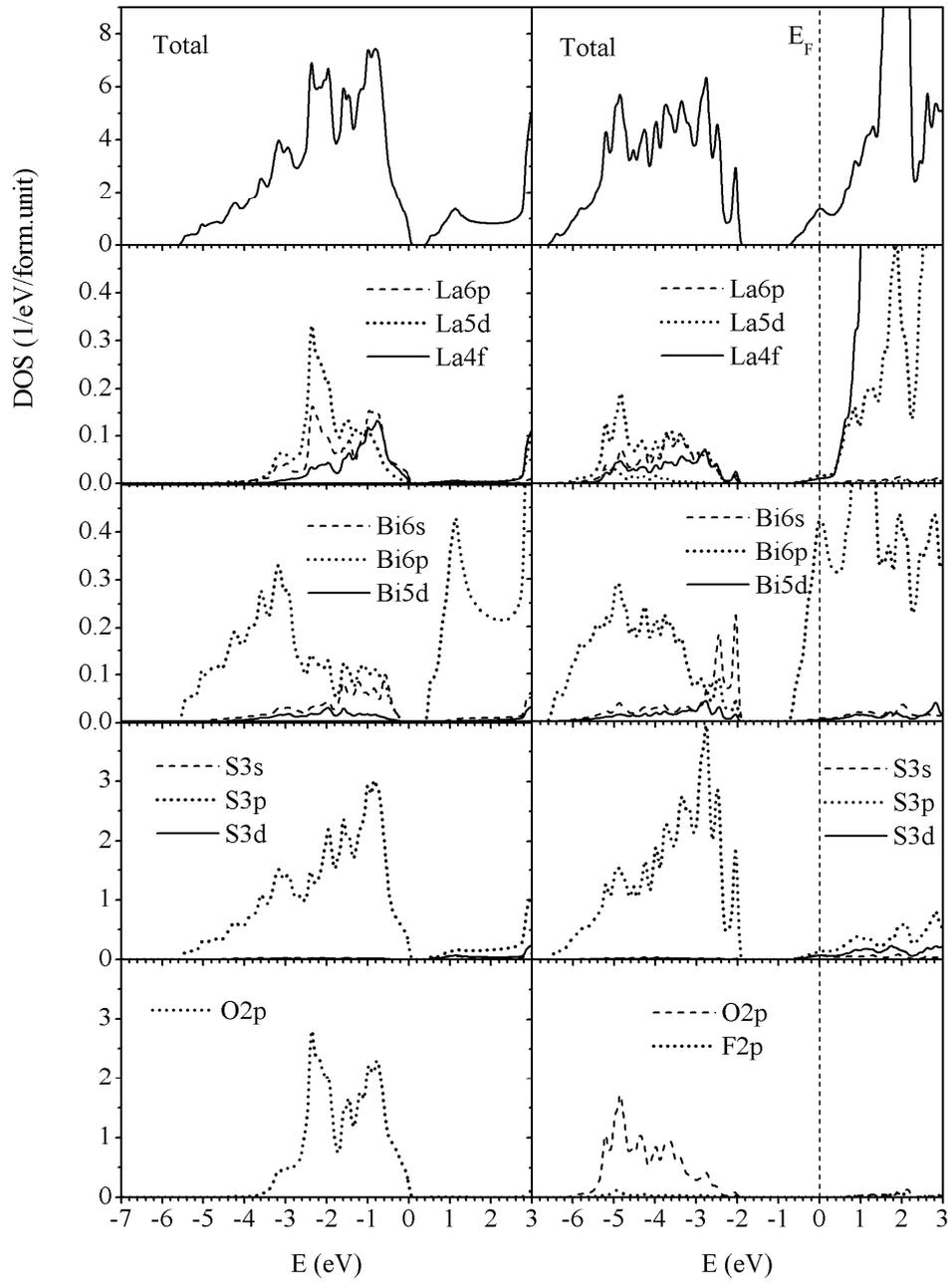

Fig. 3. Total and partial densities of states for LaOBiS$_2$ (*on the left*) and LaO$_{0.5}$F$_{0.5}$BiS$_2$ (*on the right*).

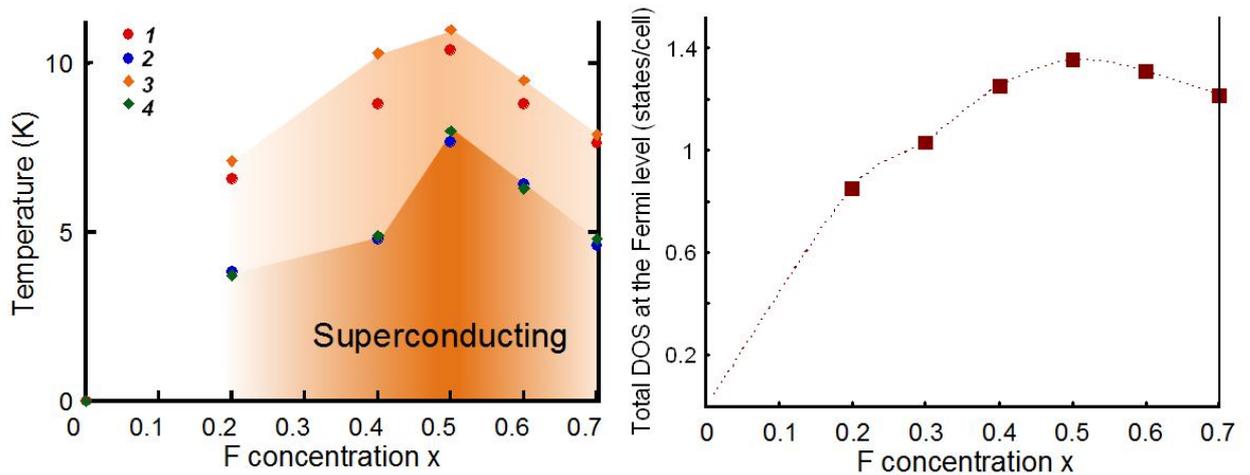

Fig. 4. The experimental [18] superconducting phase diagram of $LaO_{1-x}F_xBiS_2$ (where $1$ is the onset temperature ($T_C^{onset}$); $2$ is the zero-resistivity temperature ($T_C^{zero}$); $3$ is $T_C^{mag}$ defined as the magnetization onset temperature, and $4$ is $T_C^{irr}$ defined as the starting temperature of bifurcation between $\chi_{ZFC}$ and $\chi_{FC}$) and the calculated dependence of the total DOSs at the Fermi level on stoichiometry index (x) for this system.

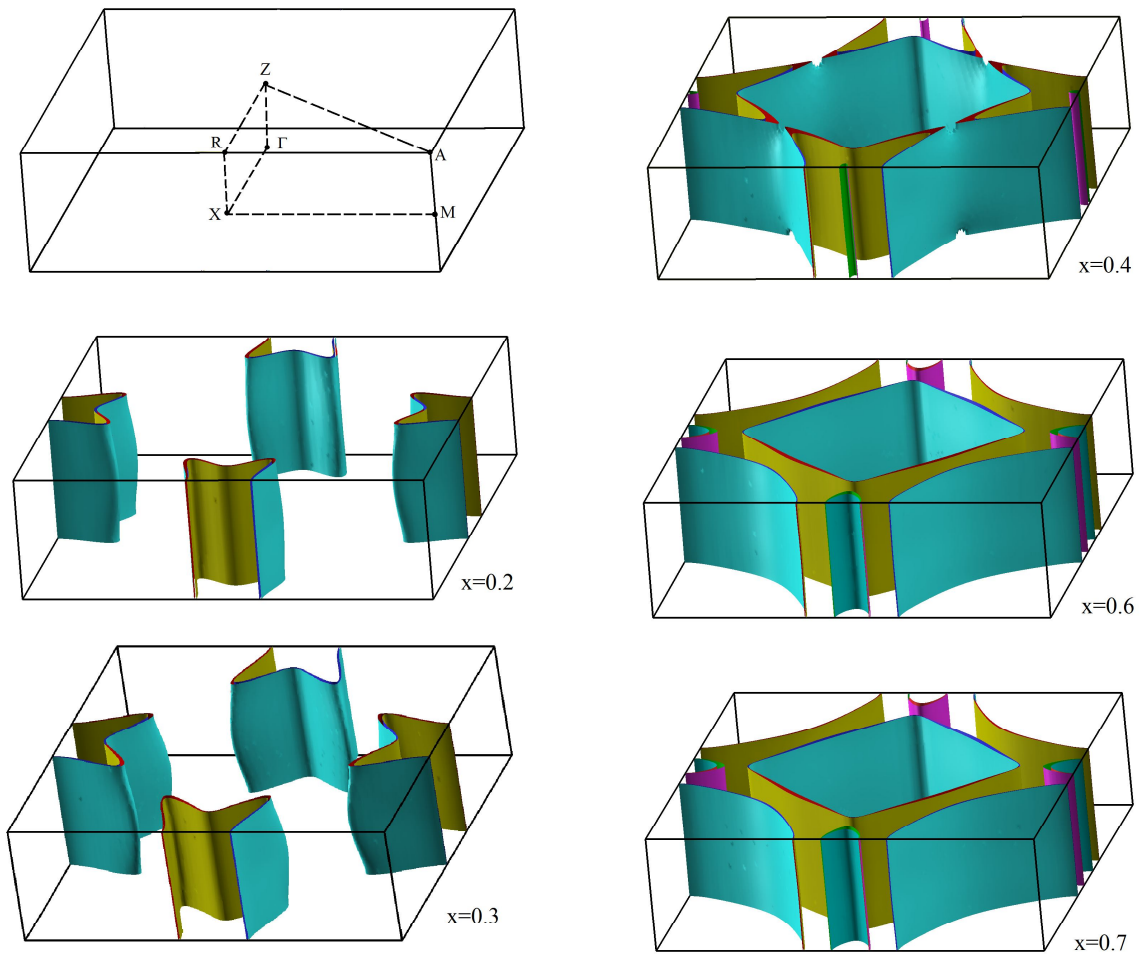

Fig. 5. Brillouin zone and the Fermi surfaces for $LaO_{1-x}F_xBiS_2$ at various doping levels x.